\pdfoutput=1
\RequirePackage{ifpdf}
\ifpdf 
\documentclass[pdftex]{sigma}
\else
\documentclass{sigma}
\fi

\numberwithin{equation}{section}

\newtheorem{Theorem}{Theorem}[section]
\newtheorem*{Theorem*}{Theorem}
\newtheorem{Corollary}[Theorem]{Corollary}

\newtheorem{Proposition}[Theorem]{Proposition}
 { \theoremstyle{definition}
\newtheorem{Definition}[Theorem]{Definition}

\newtheorem{Example}[Theorem]{Example}
\newtheorem{Remark}[Theorem]{Remark} }

\begin{document}
\allowdisplaybreaks

\newcommand{\arXivNumber}{2208.12690}

\renewcommand{\PaperNumber}{062}

\FirstPageHeading

\ShortArticleName{Separation of Variables and Superintegrability on Riemannian Coverings}

\ArticleName{Separation of Variables and Superintegrability\\ on Riemannian Coverings}

\Author{Claudia Maria CHANU~$^{\rm a}$ and Giovanni RASTELLI~$^{\rm b}$}

\AuthorNameForHeading{C.M.~Chanu and G.~Rastelli}

\Address{$^{\rm a)}$~Dipartimento di Scienze Umane e Sociali, Universit\`{a} della Valle d'Aosta, Italy}
\EmailD{\href{mailto:c.chanu@univda.it}{c.chanu@univda.it}}

\Address{$^{\rm b)}$~Dipartimento di Matematica, Universit\`{a} di Torino, Italy}
\EmailD{\href{mailto:giovanni.rastelli@unito.it}{giovanni.rastelli@unito.it}}

\ArticleDates{Received January 11, 2023, in final form August 23, 2023; Published online September 03, 2023}

\Abstract{We introduce St\"ackel separable coordinates on the covering manifolds $M_k$, where~$k$ is a rational parameter, of certain constant-curvature Riemannian manifolds with the structure of warped manifold. These covering manifolds appear implicitly in literature as connected with superintegrable systems with polynomial in the momenta first integrals of arbitrarily high degree, such as the Tremblay--Turbiner--Winternitz system. We study here for the first time multiseparability and superintegrability of natural Hamiltonian systems on these manifolds and see how these properties depend on the parameter $k$.}

\Keywords{Riemannian coverings; integrable systems; separable coordinates}

\Classification{70H06; 58J60}

\section{Introduction}

Several times in the literature~\cite{TTWcdr,CDRPW,MPW,PW,TTW} natural Hamiltonians
of the form
 \begin{equation*}
H=\frac 12 p_r^2 +\frac 1{2r^2}\big( p_\phi^2 +V(\sin(h \phi), \cos(h\phi))\big) +F(r)
 \end{equation*}
are considered as Hamiltonian in polar coordinates defined on the Euclidean plane for any real non-zero value of $h$.
In some of these papers~\cite{TTWcdr,CDRPW} these Hamiltonians are
rewritten by the rescaling $\psi=h\phi$ as
\begin{equation}\label{ham}
 H=\frac 12 p_r^2 +\frac {h^2} {2 r^2}\bigg( p_\psi^2 +\frac{1}{h^2}V(\sin(\psi), \cos(\psi))\bigg) +F(r),
\end{equation}
and interpreted as Hamiltonians on manifold with a warped product metric. These Hamiltonians have been considered recently in several articles devoted to the study of the superintegrability of Hamiltonian systems~\cite{TTWcdr,CDRPW,MPW,PW,TTW}. However, in that literature, the global structure of these manifolds has not been considered into detail.
In this paper, we study the metric
of the rescaled Hamiltonian~(\ref{ham})
as a metric on a Riemannian covering of the Euclidean plane and enlighten the differences of this alternative approach.
Thus, we are lead to study a class of Riemannian manifolds whose metric tensor depends on a integer, rational or even real parameter $k$. The aim of our work is to start a deeper analysis of these manifolds. We see how~-- in most of the cases we are going to consider~-- they can be understood as Riemannian coverings of the manifold obtained for $k=1$: the Euclidean plane. In this way, we see from a geometric view point that problems of global definition of Hamiltonians and of their first integrals can arise for non integer values of $k$, problems not yet noticed in literature.

Moreover, we show, apparently for the first time, how separable coordinates for the Hamilton--Jacobi equation can be introduced in the Riemannian coverings and we study them in several examples.

Our main result is about the superintegrability of the Tremblay--Turbiner--Winternitz, the Post--Winternitz and, consequently, of the harmonic oscillator and Kepler--Coulomb systems. The introduction of Riemannian coverings allows us to understand in a better way issues of global definition for the first integrals of high degree of those systems determined in~\cite{PW,TTW}, and allow us to show the global definition of those determined via the extension procedure in~\cite{TTWcdr,CDRPW}. In particular, we show that the Kepler--Coulomb system on Riemannian coverings does not admit as first integral the standard Laplace vector, instead, it is replaced by a first integral of degree depending on $k$.

We recall that an $n$-dimensional Hamiltonian system is (maximally) \emph{superintegrable} when it admits $2n-1$ functionally independent and globally defined first integrals; if the first integrals are all quadratic in the momenta, the system is called quadratically superintegrable.

Riemannian manifolds depending on a parameter that can be considered as Riemannian coverings appear in recent studies describing accelerated black holes~\cite{intro}. Metrics of this type do appear also in orbifold theory, see, for example,~\cite{coop}, where it is made use of the discrete, dihedral symmetries introduced by the parameter in the metric. Another application is made in~\cite{DG} to represent circularly symmetric $N$-vortex solutions of elliptic sinh-Gordon and Tzitzeica equation.

In Section~\ref{sec2}, we introduce the main two-dimensional Riemannian manifold we are considering; we give a definition of Riemannian coverings and we provide a summary of the geometric theory of separation of variables for the Hamilton--Jacobi equation. In Section~\ref{sec3}, we analyze the map determining the Riemannian coverings and its representation on the Euclidean plane. In Section~\ref{sec4}, we introduce and study separable coordinates for the Hamilton--Jacobi equation on Riemannian coverings of the Euclidean plane, with examples in other two-dimensional manifolds. In Section~\ref{sec5}, we discuss the separability and superintegrability of the harmonic oscillator, the Kepler--Coulomb, the Tremblay--Turbiner--Winternitz and the Post--Winternitz systems, when considered as systems on covering manifolds, as done in some literature, and we make a comparison with the superintegrability of the same systems when they are not defined on covering manifolds. In Section~\ref{sec6}, we resume and discuss the main results of the paper.

	\section[Separable coordinates on Riemannian coverings: introduction and definitions]{Separable coordinates on Riemannian coverings: \\ introduction and definitions}\label{sec2}

\begin{Definition}[\cite{BenentiSystems}]
 An orthogonal coordinate system $ \big(q^h\big)$ on an~$n$-dimensional Riemannian manifold $ (M,g)$ is called \emph{separable} for a natural Hamiltonian
 \[
 H= \frac 12 g^{jj}\big(q^h\big) p_j^2 +V\big(q^h\big), \qquad h,j =1,\ldots, n,
 \]
 if
the corresponding Hamilton--Jacobi equation $H \big(q^j,\partial_{q^j} S\big(q^h\big) \big)=E$ admits a solution in additive separated form $S=S_1\big(q^1,c_h\big)+\dots + S_n(q^n,c_h)$ depending on $n$ parameters satisfying the completeness condition $\det \big(\partial^2_{q^h c_j} S\big)\neq 0$.
\end{Definition}

Even if separability is a local feature, nevertheless the existence (and the determination) of separable coordinates for natural Hamiltonians can be characterised through Riemannian geometric objects which are global or almost global defined on the Riemannian manifold $M$.

\begin{Theorem}[\cite{BenentiSystems}]
 A natural Hamiltonian on $(M,g)$ admits orthogonal separable coordinates if and only if there exists a Killing $2$-tensor $K$ for $g$ with simple eigenvalues and normal eigenvectors, satisfying the compatibility condition
 \[
 {\rm d}K{\rm d}V=0,
 \] where $K$ is interpreted as linear operator on $1$-forms, i.e., as a $(1,1)$-tensor. The separable coordinates hypersurfaces are orthogonal to the eigenvectors of $K$.
 \end{Theorem}
 We recall that a $2$-contravariant symmetric tensor ia a Killing tensor if and only if $K$ corresponds to a quadratic first integral of the geodesic Hamiltonian; in terms of Poisson bracket on~$T^*M$ we have
 \[
 \big\{K^{ij} p_ip_j, g^{ij}p_ip_j\big\}=0.
 \]
Separability for a natural Hamiltonian implies Liouville integrability, that is the existence of $n$ independent first integrals pairwise in involution:
\begin{Corollary}
 If a natural Hamiltonian admits a separable orthogonal coordinate system then it has other $n-1 $ independent pairwise commuting quadratics in the momenta integrals of motion. The converse it is true if the quadratic first integrals share the same eigenvectors as linear operators.
\end{Corollary}
Moreover, the hypothesis of normality of the eigenvectors of a Killing tensor $K$ (i.e., the fact that eigenvectors of $K$ are orthogonally integrable) is automatically satisfied on any $2$-dimensional Riemannian manifold. The eigenvalues also allow us to determine the separable coordinates:
\begin{Proposition}[see~\cite{CReig} for the general $n$-dimensional case]\label{Peig}
On a $2$-dimensional Riemannian manifold, the level sets of the $($non-constants$)$ eigenvalues of a compatible Killing $2$-tensor define the separable hypersurfaces.
\end{Proposition}

So the basic necessary objects to construct separable coordinates are Killing $2$-tensors with some additional properties. Moreover, we have that

\begin{Proposition}[\cite{KL}]\label{P4}
On a constant curvature $n$-dimensional Riemannian manifold, the vector space of Killing $2$-tensors has the maximal possible dimension \[ n(n+1)^2(n+2)/12 \] and it is spanned by symmetric products of Killing vectors, that is infinitesimal symmetries of the metric.
\end{Proposition}

 \begin{Remark}
 Being of constant curvature is a local property of a Riemannian manifold, thus the globality of the tensors also depends on the topological aspects.
\end{Remark}
\begin{Remark}
 In the Euclidean plane, all the possible orthogonal separable coordinate systems are one of the well known Cartesian, polar, parabolic and elliptic-hyperbolic coordinates, each of them generated by a different $K$.
\end{Remark}
	Let us consider a two-dimensional Riemannian manifold $M_k$ endowed with the metric
	\begin{equation}\label{wm}
	g_k={\rm d}r^2+k^2r^2 \,{\rm d}\phi^2,
	\end{equation}
	where
	 \begin{equation*}
	0<r, \qquad 0\leq \phi <2\pi, \qquad k\in \mathbb R^+.
	 \end{equation*}
	The Riemannian manifold $M_k$ can be considered as the warped product of a circle, parametrized by $\phi$ and a (open) half line, parametrized by $r$.
	Hence, the metric is globally defined on a cylinder or a cone or on
	a punctured plane (i.e., the open set $\mathbb R^2\setminus \{(0,0)\}$).

\begin{Proposition}\quad
 \begin{enumerate}\itemsep=0pt
	\item[$(i)$] For any $k$, the metric $g_k$ is flat and coincides locally with the metric of the Euclidean plane~$\mathbb{E}^2$ $($for $k=1$ this fact holds globally$)$. The metric $g_k$ becomes the metric of the Euclidean plane after the transformation
 \begin{equation}
 \Pi\colon \ M_k \rightarrow \mathbb{E}^2=M_1,\qquad (r,\phi)\mapsto (r, \Phi=k\phi), \label{PI}
 \end{equation}
 where the angle $\Phi$ now runs between $0$ and $2k\pi$.
 \item[$(ii)$] the coordinates $(r,\phi)$ are not the standard polar coordinates of the Euclidean plane.
 \end{enumerate}
	\end{Proposition}
\begin{proof}
Item $(i)$ is straightforward.
 An easy way to show $(ii)$ is to apply Proposition \ref{Peig} and determine the separable coordinate hypersurfaces, which are curves in two-dimensional case.
	These curves are the level sets of the eigenvalues $\rho_1$, $\rho_2$ of the Killing tensor, provided they are both distinct and non-constant. By using a computer-algebra software, for example, \textsc{Maple}, we can easily compute the symmetric Killing $2$-tensors of $g_k$, which will depend on trigonometric functions of $k\phi$, and we choose one of them, $K$, for example, associated with elliptic-hyperbolic coordinates of the Euclidean plane when $k=1$, and we draw the level sets of eigenvalues passing through a given point of the manifold. By plotting these curves in the plane with respect to standard polar coordinates $(r,\phi)$, the orthogonal separable coordinates determined by the
 eigenvalues of $K$, for $k=1, 1/2, 2$ are shown in the three images of Figure~\ref{f1}.
		
\begin{figure}[t]
	\centering
		\includegraphics[width=3.4cm]{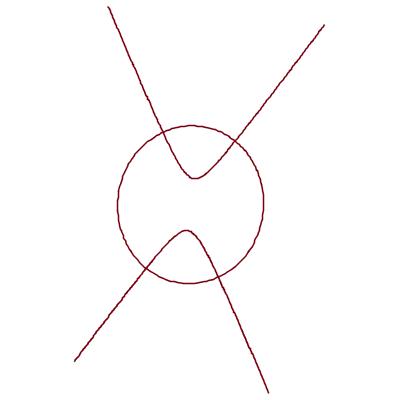}
					\includegraphics[width=3.4cm]{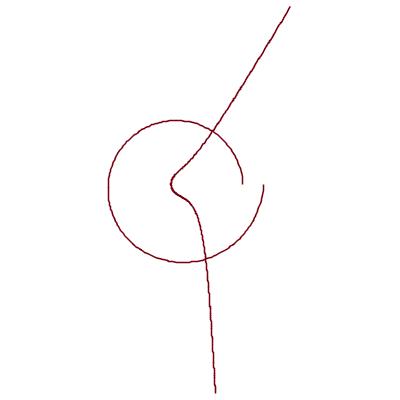}
					\includegraphics[width=3.4cm]{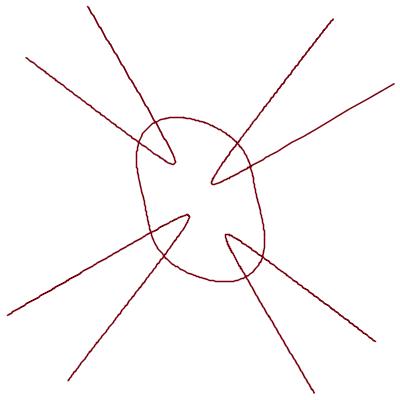}
		\caption{``Orthogonal" coordinates for $k=1$, $k=1/2$, $k=2$.}\label{f1}
\end{figure}

A close inspection shows that only for $k=1$ these curves intersect each other orthogonally, moreover, for $k=1/2$ the curve expected to be closed, actually is not. We conclude that the map $(r,\phi)\mapsto (x=r\cos \phi, y=r\sin \phi)\in \mathbb{E}_2$ is not conformal for $g_k$ if $k\neq 1$, since it maps orthogonal coordinate webs with respect to $g_k$ into nonorthogonal webs for the Euclidean metric. Indeed, it is straightforward to see that $g_k$ is conformal to the Euclidean metric if and only if $k^2=1$. Moreover, we see that the separable coordinates assume shapes unexpected for separable coordinates in Euclidean spaces, see the figure with $k=2$, for example, with problems of global definition for non-integer values of $k$, for example, $k=1/2$.
\end{proof}

Clearly, the theory of separation of variables cannot be applied to metrics such as $g_k$ without a deeper understanding of the Riemannian manifold $M_k$, $g_k$. The global definition of the Killing tensors associated with the coordinates plays here a determinant role. In this case the components of the tensor determining the separable coordinates include trigonometric functions of $k\phi$ (see Proposition \ref{P11} below). This means that the metric and the tensors are assumed to live on a covering manifold of the Euclidean plane, instead of on the Euclidean plane itself.
Indeed, the Riemannian manifold $(M_k,g_k)$ is the Riemannian manifold obtained by pulling-back the Riemannian metric of $M_1$ on a covering of $M_1$ determined by the map $\Pi$ given in (\ref{PI}).

We recall the definition of Riemannian coverings as introduced in~\cite{IC}.
\begin{Definition} If $M$ and $N$ are connected topological manifolds, we say that a map $\psi\colon M\to N$ is a \emph{covering} if every $p \in N$ has a connected open neighborhood $U$ such that $\psi$ maps each component of $\psi^{-1}[U]$ homeomorphically onto $U$.
If $M$ and $N$ are also differentiable manifolds, then $\psi$ is a \emph{differentiable covering} if $\psi $ is differentiable of maximum rank on $M$.
If $M$ and $N$ are also Riemannian manifolds, then $\psi$ is a \emph{Riemannian covering} if $\psi $ is differentiable of maximum rank and a local isometry of $M$ on $N$ (i.e., the metric on $M$ is the pull-back via $\psi$ of the metric on $N$ wherever the Jacobian linear map $\psi_*\colon TM \to TN$ is one-to-one).
\end{Definition}

\begin{Proposition}\label{Proposition4}
 For $k$ integer and $k \geq 1$, $\Pi$ is a
Riemannian covering map.
\end{Proposition}

\begin{proof}
Since $M_k$ is (path) connected, $\Pi$ is surjective and continuous and for each point
$p\in M_1$ there exist a connected neighborhood $U$ of $p$, such that $\Pi^{-1}(U)$ is mapped diffeomorphically
by~$\Pi$ on $U$, since $\Phi$ spans $k$ times the plane. Then, since $g_k$ is the pull-back of $g_1$, $\Pi$ is a local isometry, and
$\Pi$ is a Riemannian covering map.
\end{proof}

\begin{Remark}
Whenever $k$ is not integer, $M_k$ fails to be a covering of $M_1$, because not all the points in $M_k$ have neighborhoods that can be projected on $M_1$ consistently with their topology. Indeed, in these cases the projections on $M_1$ of neighborhoods of points with $\Phi=0$ and $\Phi=2k\pi$ cannot be made to coincide. For simplicity, however, we call coverings also the $M_k$ with $k$ not integer, and we leave to the nature of $k$ the distinction between the true coverings and those that are not such.
\end{Remark}

Global properties of geometric objects on $M_k$ could differ from those of the corresponding objects of $M_1$.

Riemannian metrics of the form (\ref{wm}) have been recently considered in several articles concerning superintegrable systems with polynomial first integrals of arbitrarily high degree~\cite{MPW, TTW}. In some of these articles, although essentially correct, it appears nevertheless some ambiguity on the identification of the manifold where (\ref{wm}) is defined, often simply denoted as Euclidean plane (as well as with the obvious modifications of the metric, sphere, pseudosphere, etc.), even if in some cases the actual manifold is a Riemannian covering of the above-mentioned manifold. In particular, in none of these articles there is any attempt to study separable coordinates for the metric $g_k$ on Riemannian coverings, and many of the problems we are facing here do not arise there.

\begin{Remark} In several articles, see~\cite{TTWcdr} and references therein, it has been developed a procedure, called extension procedure, for constructing first integrals of arbitrarily high degree of certain warped natural Hamiltonians, including those with metric (\ref{wm}), when $k$ is any rational number. The procedure does not make use of the map $\Pi$ in an essential way, i.e., the map is introduced only to build certain examples, such as the three-body Calogero system, or the TTW system. In these cases, whenever $k$ is not integer, problems of non-global definition can arise according to the topology of the manifold under consideration, as we will see later in this paper. These problems were not always taken in account in the early articles about the extension procedure. It must be remarked that all the warped Hamiltonians admitting the extension procedure described in~\cite{TTWcdr, CDRPW}, including those associated with the metric (\ref{wm}), are globally well defined for any rational $k$, as well as their first integrals originated by the procedure, since the definition of the Hamiltonians and the construction of their first integrals associated with $k$ never make use of the covering map $\Phi$.
\end{Remark}

\section[The covering map Pi]{The covering map $\boldsymbol{\Pi}$}\label{sec3}
	
	The deformed orthogonal coordinates of Figure~\ref{f1} arise from the fact that in these drawings the whole covering spaces are fitted into the Euclidean plane, then they are shrunk, for $k>1$ , or stretched, for $k<1$, under the action of $\Pi$. We see below the details of this representation. We consider only real, positive values of $k$. Let us recall that $(M_1,g_1)$ is the Euclidean plane, with polar coordinates $(r,\Phi)$, where the only, determinant, difference between these coordinates and the standard ones is that
	 \begin{equation*}
	0\leq \Phi<2k\pi.
	 \end{equation*}
Recall that $M_k$ is the Riemannian manifold obtained by pulling-back the Riemannian metric of~$M_1$ on a covering of~$M_1$ determined by the map $\Pi\colon (r,\phi)\rightarrow (r,\Phi)$.

	We can write the standard coordinate change into Cartesian coordinates $(x,y)$ as
	\begin{gather*}
	x=r \cos \Phi,\qquad
	y=r \sin \Phi,
	\end{gather*}
	with inverse transformation
	\begin{gather*}
	r^2=x^2+y^2,\qquad
	\Phi=\arctan \frac yx.
	\end{gather*}
	Since $\Phi=k\phi$, we have
	 \begin{equation*}
	\phi=\frac 1k \arctan \frac yx.
	 \end{equation*}
It follows
 \begin{Proposition}
	The transformation $(r,\Phi)\leftrightarrow (x,y)$ is well defined and biunivocal only when
\begin{equation}\label{sec}
m\frac{2\pi}k \leq \phi	< (m+1) \frac{2\pi}k, \qquad m \in \mathbb N, \quad (m+1)\leq k,
\end{equation}
and, if $k$ is not integer, in the incomplete sector
 \begin{equation*}
([k]-1)\frac{2\pi}{[k]}\leq \phi < k\frac{2\pi}{[k]}, \qquad [k]\geq 1,
 \end{equation*}
to its image in $M_1$, where $[k]$ denotes the integer part of $k$.
\end{Proposition}

\begin{proof} The $[k]$ sectors, plus the incomplete sector, make a partition of the domain of $\phi$ where each sector is mapped in the Euclidean plane, and in a portion of it for the incomplete sector. \end{proof}

Due to the isometric correspondence between sectors and Euclidean planes seen above, we see that
\begin{Proposition}
The Killing tensors of $g_k$ are the same Killing tensors of the Euclidean plane in each sector, as defined by~\eqref{sec}
\end{Proposition}

\begin{figure}
	\centering

	\includegraphics[width=3.5cm]{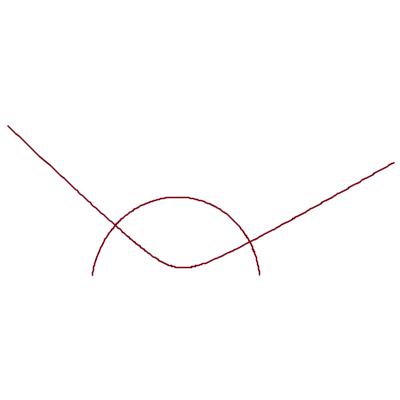}
	\includegraphics[width=3.5cm]{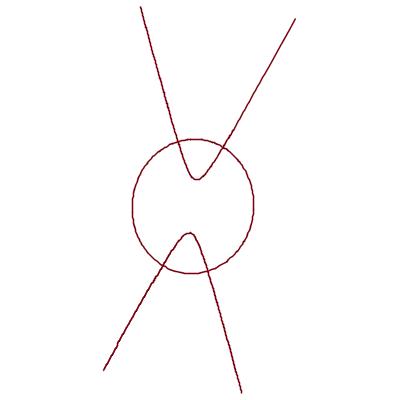}
	
	\caption{Cases $k=1/2$ and $k=2$ of Figure~\ref{f1} plotted in $(r,\Phi)$ coordinates, with $0\leq \Phi < \pi$ and $0\leq \Phi < 2\pi$ respectively.}
	\label{f4}
\end{figure}

\begin{proof} We call $\Phi_m$ the map $\Phi=k \phi$ restricted to the $m$-th sector of the covering. It follows from the computation of the Killing tensors in coordinates $(r,\Phi_m)$ for each sector. All these transformations, understood as transformations
$(r,\Phi_m)\leftrightarrow (x,y)$, map each sector (\ref{sec}) spanned by coordinates $(r,\Phi_m)$, $(x,y)$, into the Euclidean plane, and a portion of the Euclidean plane into itself for the incomplete sector, in a conformal way, as shown in Figure~\ref{f4}.
\end{proof}

\begin{Corollary} The separable coordinates for $g_k$ are the union of the separable coordinates in the $[k]$ Euclidean planes, namely Cartesian, parabolic, polar and elliptic-hyperbolic, plus a part of those in the portion of Euclidean plane corresponding to the incomplete sector.
\end{Corollary}

Remark that, in this way, each one of the $[k]$ sectors where $\Phi_m$ varies between multiples of~$2\pi$ is a whole Euclidean plane (up to the singular point $r=0$).

\begin{figure}[t]
	\centering
	
	\includegraphics[width=12.5cm]{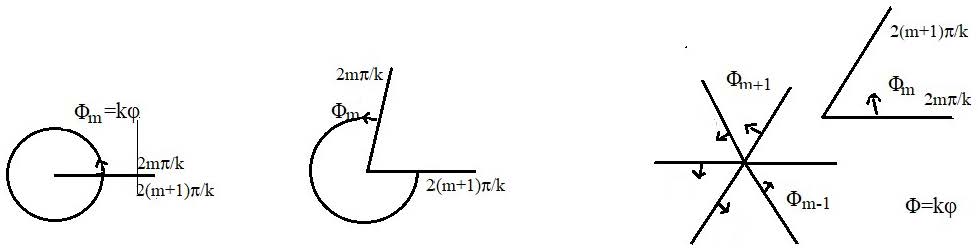}
		\caption{The map $(r,\phi)$ as union of the maps $(r,\Phi)$.}
		\label{f2}
\end{figure}

Under the map $\Pi$, each one of the $[k]$ planes is cut along the half-line $\Phi=2m\pi/k$, then shrunk to fit the sector of width $2\pi/[k]$ in $(r,\phi)$ coordinates, with its borders glued to those of the neighbouring sectors (see Figure~\ref{f2}). When $k$ is not integer, one of the sectors covers only a~portion of the plane.

\begin{figure}[t]
	\centering
	
	\includegraphics[width=12.5cm]{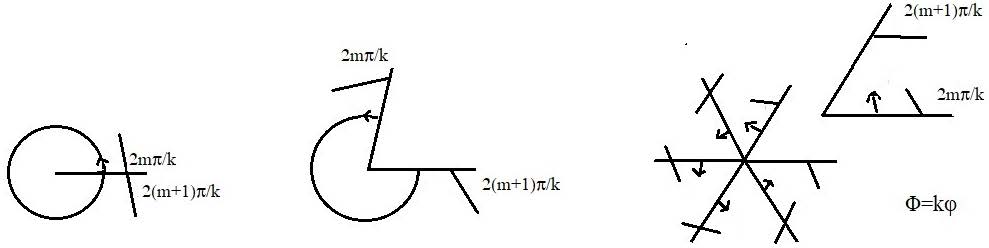}
	\caption{Curves intersecting the cutting line are reconnected.}
	\label{f3}
\end{figure}

Despite its global non-conformality, the map $(r,\phi)$ illustrated in Figure~\ref{f2} shows well as, for integer $k$, the globality of the separable curves, as well as of any global coordinate system on the planes $(r,\Phi)$, is preserved. Indeed, any curve intersecting (smoothly) the cutting line $\Phi=2m\pi/k$ becomes a segment of curve whose end points, separated by the cut, are glued consistently with the end points of the copies of the same curve in the neighbouring sectors, building a continuous (smooth) curve (see Figure~\ref{f3}).

The Euclidean planes corresponding to each sector are glued together according to Figure~\ref{f2} and the corresponding separable coordinates, cut and glued together consequently, result for example as in Figure~\ref{f1}, determining, for $k$ integer, global orthogonal separable coordinates on the manifold $M_k$. For non integer values of $k$, as for $k=1/2$ in Figure~\ref{f1}, the resulting curves on $M_k$ are not, in general, globally defined.

If we plot the cases $k=1/2$ and $k=2$ of Figure~\ref{f1} in polar coordinates $(r,\Phi)$, with $0\leq \Phi < \pi$ and $0\leq \Phi < 2\pi$ respectively, we obtain the familiar separable coordinates of $\mathbb E^2$, as done in Figure~\ref{f4}, and we can appreciate the local isometry of the map $\Pi$.

\begin{figure}
	\centering
	
	\includegraphics[width=3.5cm]{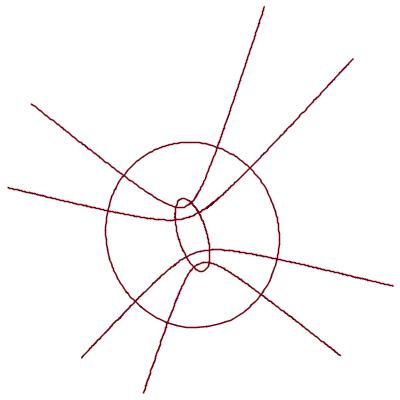}
	\includegraphics[width=3.5cm]{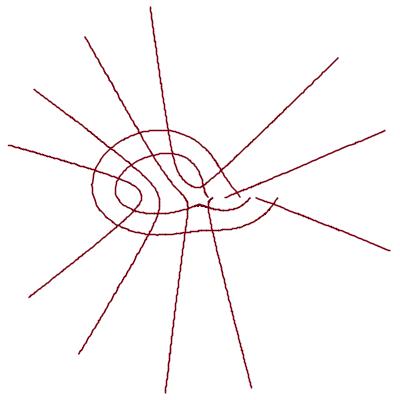}
	\includegraphics[width=3.5cm]{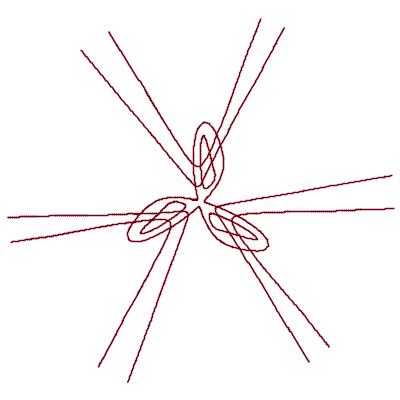}
	
	\caption{Other separable coordinates, generated by the same Killing tensor parametrized by $k$, for $k=1$, $k=3/2$, $k=3$, plotted in coordinates $(r,\phi)$.}
	\label{f5}
\end{figure}

It is important to stress the fact that the global separable coordinate web of $(M_k,g_k)$ generated by $K$ assumes the form obtained by plotting the eigenvalues of $K$ in coordinates $(r,\phi)$, keeping in mind that this map is not conformal, so we cannot expect the preservation of orthogonality of the separable web. In Figure~\ref{f5}, we show other examples for a different $K$ and different values of $k$. We can now understand the dihedral symmetry introduced by $\Pi$ in the plane as the result of squeezing several copies of the plane around the origin, and the unusual shape of the separable coordinates in each $M_k$ as the result of the same squeezing of several copies of the plane, together with the corresponding separable coordinate systems, around the origin. We see below that for other manifolds, the sphere $\mathbb S^2$ for example, the behaviour of the covering map is more immediately understandable.
\begin{Remark}
When $k<1$, the whole manifold $M_k$ covers, via $\Pi$, only a portion of the Euclidean plane $M_1$, corresponding to an incomplete sector as introduced above. Therefore, the separable coordinates pulled back on $M_k$ are in this case necessarily incomplete, as shown in Figures \ref{f1} and \ref{f4} for $k=1/2$.
\end{Remark}

\section[Separable coordinates on Riemannian coverings: Killing vectors and tensors]{Separable coordinates on Riemannian coverings: \\ Killing vectors and tensors}\label{sec4}

We are going to study, apparently for the first time, on the covering manifolds $M_k$ the separability structures pulled-back from those of the base of the covering $M_1$.
\begin{Proposition} Any Killing vector $V$ for~\eqref{wm} can be written as linear combination with constant real coefficients of the three vectors
\begin{equation*}
 V_1=\frac 1{r}\partial_\phi, \qquad V_2=- k \cos k\phi \partial_r+\frac {\sin k\phi}{r^2}\partial_\phi, \qquad V_3=-k \sin k\phi \partial_r+\frac {\cos k\phi}{r^2}\partial_\phi,
\end{equation*}
so that
\begin{equation}\label{kv}
V=k(a_3 \sin k \phi-a_2 \cos k \phi) \partial_r+\frac {a_3 \cos k \phi+a_2 \sin k \phi +a_1 r}{ r^2}\partial_\phi,
\end{equation}
where $(a_i)$ are real constants.
\end{Proposition}

\begin{proof} By direct integration of the Killing equations. \end{proof}

\begin{Proposition}\label{P11}
Any symmetric Killing two-tensor $K$ for~\eqref{wm} can be written, in contravariant form, as linear combination with real constant coefficients of
\begin{gather}
 k^2 \partial_r \odot \partial_r+ \frac 1{r^2}\partial_\phi \odot \partial_\phi, \nonumber\\
 \partial_\phi \odot \partial_\phi, \label{phph} \\
 k \sin k\phi \partial_r \odot \partial_\phi+\frac 1{r}\cos k\phi \partial_\phi \odot \partial_\phi,\nonumber\\
 -k \cos k\phi \partial_r \odot \partial_\phi+\frac 1{r}\sin k\phi \partial_\phi \odot \partial_\phi,\nonumber\\
 -k^2\sin \, 2k \phi \partial_r \odot \partial_r- \frac {2k}{r}\cos \, 2k\phi \partial_r \odot \partial_\phi+ \frac 1{r^2}\sin \, 2k\phi \partial_\phi \odot \partial_\phi, \label{16}\\
 -k^2\cos \, 2k \phi \partial_r \odot \partial_r+ \frac {2k}r\sin \, 2k\phi \partial_r \odot \partial_\phi+ \frac 1{r^2} \cos 2k\phi \partial_\phi \odot \partial_\phi,\label{17}
\end{gather}
so that
\begin{gather}
	K=k^2(-b_5 \sin 2k\phi-b_6\cos 2k\phi +b_4)\partial_r \odot \partial_r \nonumber
\\ \hphantom{K=}{} +k\, \frac{b_3r\sin k\phi-b_2r\cos k\phi+2b_6\sin 2k\phi-2b_5\cos 2k\phi}{r} \partial_r \odot \partial _\phi \nonumber
\\ \hphantom{K=}{} +\frac{b_2r \sin k\phi+b_3r\cos k\phi+b_1r^2+b_5\sin 2k\phi+b_6\cos 2k\phi+b_4}{r^2} \partial_\phi \odot \partial_\phi, \label{kt}
\end{gather}
where $(b_i)$ are real constants and $\odot$ denotes the symmetric product of tensors.
\end{Proposition}

\begin{proof} By Proposition \ref{P4}, since the metric is flat, Killing tensors are all the linear combinations with constant coefficients of the symmetric products of the Killing vectors. \end{proof}

These tensors are well defined on the covering $M_k$ of $M_1$ if and only if their components have a period in $\phi$ which is integer multiple of $2\pi$. This certainly always happens when $k$ is an integer, but not always when $k$ is rational, as we are going to see below. Therefore,

\begin{Proposition} The dimension of the space of the global Killing tensors on $M_k$ is a function of the order of the tensor and of $k$, even if $M_k$ is a flat manifold and, therefore, the dimension of the $($local$)$ Killing tensor space is maximal $($see Proposition~$\ref{P4}$$)$.
\end{Proposition}

As instance, we see from (\ref{kv}) that for $k=\frac 12$ the dimension of the space of the Killing vectors is one instead of three, the dimension of the space for any non null integer value of $k$, since the only generator globally defined is then $\frac 1r \partial_\phi$. From (\ref{kt}), it follows that

\begin{Proposition} For $k$ integer greater than zero, the space of the globally defined symmetric Killing vectors is described by~\eqref{kv} and has dimension six.
For any other non-zero real value of~$k$, it has dimension one and it is generated by
\begin{equation*}
 \frac 1r \partial_\phi.
\end{equation*}
\end{Proposition}

\begin{Proposition}\label{10} For $k$ integer greater than zero, the space of the globally defined symmetric Killing $2$-tensors is described by~\eqref{kt} and has dimension six.
For $k=\frac q2$, where $q$ is any odd integer, it has dimension four, and it is generated by the linear combinations of the contravariant metric tensor
\begin{equation*}
g_k=\partial_r \odot \partial_r+ \frac 1{k^2r^2}\partial_\phi \odot \partial_\phi,
\end{equation*}
and \eqref{phph}, \eqref{16}, \eqref{17}.
 For any other real non-zero value of $k$, it has dimension two and it is generated by the linear combinations of the metric tensor itself and
 \eqref{phph}.
\end{Proposition}

\begin{Example} Let us consider the separable coordinates on a circular cylinder in $\mathbb E^3$. Since the cylinder is locally flat, it admits locally the same Killing tensors of $\mathbb E^2$, then the same separable coordinate systems of $\mathbb E^2$. However, not all these tensors are globally defined on the cylinder, so that, of all the separable coordinates of $\mathbb E^2$, only those compatible with the global structure of the cylinder remain, namely, the Cartesian coordinates of $\mathbb E^2$ having a coordinate direction parallel to the axis of the cylinder. Actually, the space of Killing vectors on the cylinder is only two-dimensional, instead of three-dimensional, generated by the translation along the axis of the cylinder and by the rotation around the same axis. The space of the symmetric Killing two tensors, spanned by the linear combinations with constant coefficients of the symmetric products of the Killing vectors, is therefore of dimension three instead of six.
\end{Example}
\begin{Proposition} The coordinates $(r,\phi)$ and $(r,\Phi)$ are associated with the tensor \eqref{phph} and separable on any $M_k$.
	\end{Proposition}

 \begin{proof} The eigenvectors of (\ref{phph}) are indeed $\partial_r$ and $\partial_\phi$, and the tensor is globally defined on all~$M_k$ after Proposition~\ref{10}.\end{proof}

The following statement follows from the previous propositions.

\begin{Corollary}
The Killing tensors of 	$(M_k,g_k)$ are not, in general, Killing tensors of $(M_j,g_j)$ for $j\neq k$. Their components in coordinates $(r,\phi)$ depend on trigonometric functions of $\Phi=k\phi$.
\end{Corollary}

\begin{Remark} As far as $k$ is an integer, any trigonometric function of $k\phi$ can be reduced to a~polynomial in trigonometric functions of $\phi$ via Chebyshev polynomials. It becomes then even more evident that in this case no problem of well definition of Killing vectors and tensors can arise, since everything is ultimately depending on the argument $\phi$. Whenever $k$ is a non-integer rational, instead, trigonometric functions of $k\phi$ can be expanded in function of $\phi$ only through the explicit expression of roots of algebraic equations. At this point, the expansion leads to multiple solutions and, consequently, multiple spaces of Killing vectors and tensors could be associated to the same metric (\ref{wm}). The analysis of this problem requires methods of complex geometry and algebraic geometry and we leave it for further studies.
\end{Remark}

Clearly, if we consider Hamiltonian systems of Hamiltonian
 \begin{equation*}
H=\frac 12 g_k^{ij}p_ip_j+V(r,\Phi)
 \end{equation*}
on $M_k$, then the Hamiltonian is globally well defined only if $V$ too is periodic in $\Phi=k\phi$. We assume, for simplicity, that every function is $C^\infty$ on its domain.

\begin{Remark}
We stress the fact that, even if the $M_k$ are true covering manifolds only when~$k$ is integer, one can define dynamical, Hamiltonian systems upon them for any $k\in \mathbb R\setminus\{0\}$. The $M_k$ are all diffeomorphic to $\mathbb R\times \mathbb S^1$ and all functions, vectors and tensors on $M_k$ or $T^*M_k$ depending on $\Phi$ should have a period of $2k\pi$ or its integer submultiples.
\end{Remark}

We see below several examples of Riemannian covering maps on different base manifolds and some example of the resulting separable structures on Riemannian coverings.

\subsection{Plane: parabolic coordinates}\label{PC}

The Killing tensor of $M_k$,
\begin{equation}\label{para}
K=-k{\, \cos k\phi}\, \partial _r \odot \partial_\phi+\frac{\sin k\phi}{r}\partial_\phi \odot \partial_\phi,
\end{equation}
is globally defined on the manifold for integer values of $k$ only, and it has eigenvalues
 \begin{equation*}
u=k^2\frac r{2}(\sin k \phi+1), \qquad v=k^2\frac r{2}(\sin k \phi-1),
 \end{equation*}
which can be chosen as orthogonal separable coordinates on the manifold $M_k$ with
 \begin{equation*}
r=\frac{1}{k^2}(u-v), \qquad \Phi=k\phi=\arcsin \frac{u+v}{u-v}.
 \end{equation*}
In coordinates $(u,v)$, we have
\begin{equation*}
g_k=\frac{k^4}{u-v}(u \partial_u \odot \partial_u-v \partial_v \odot \partial_v), \qquad K=\frac{k^4 uv}{u-v}(\partial_u\odot \partial_u-\partial_v\odot \partial_v),
\end{equation*}
thus, we verify that the coordinates $(u,v)$ are orthogonal and diagonalize the Killing tensor $K$.

\begin{figure}[t]
	\centering
	
	\includegraphics[width=3.5cm]{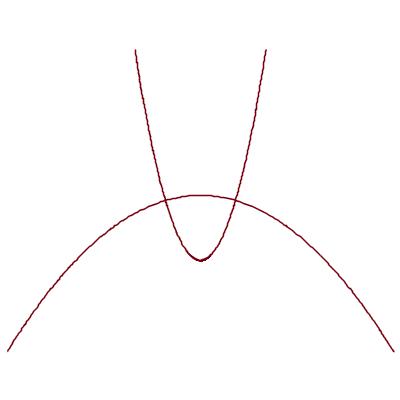}
	\includegraphics[width=3.5cm]{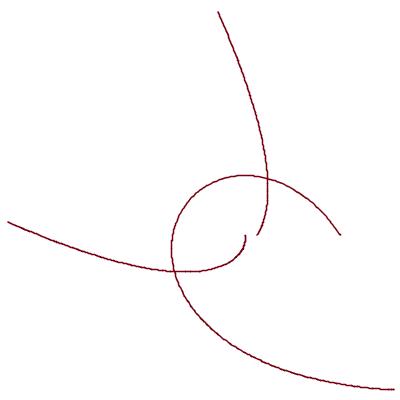}
	\includegraphics[width=3.5cm]{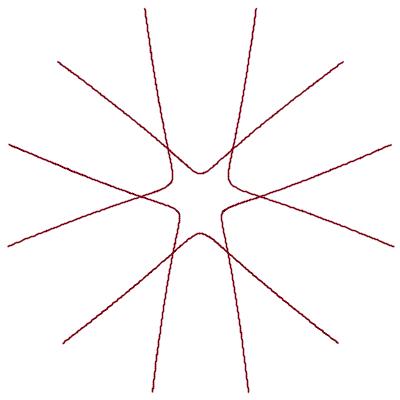}
	
	\caption{``Parabolic'' coordinates for $k=1$, $k=2/3$, $k=3$.}\label{f6}
\end{figure}

For $k=1$, the coordinates $(u,v)$ are parabolic coordinates of $\mathbb E^2$ with axis $x=0$. For $k=1$, $k=2/3$ and $k=3$ these parabolic coordinates are plotted in Figure~\ref{f6} under $(r,\phi)$ non conformal representation. A comparison with Figures~\ref{f2} and~\ref{f3} shows how the coordinate webs on $M_k$ are obtained from those of $M_1$.
	
\subsection{Sphere: spherical-conical coordinates}

We now consider the metric
 \begin{equation*}
G_k={\rm d}\theta ^2+k^2 \sin^2 \theta \, {\rm d}\phi^2
 \end{equation*}
with $k\in \mathbb R^+$, $0< \theta < \pi$, $0\leq \phi < 2\pi$. $G_k$ is locally the metric of the sphere $\mathbb S^2$ in spherical coordinates, indeed, as previously done, we introduce the new variable
 \begin{equation*}\Phi=k\phi,\qquad 0\leq \Phi < 2k\pi, \end{equation*}
and we get
 \begin{equation*}
G={\rm d}\theta^2+\sin^2\theta \, {\rm d}\Phi^2.
 \end{equation*}
Hence, we have again that each sector of width $2\pi/k$ in $\phi$ is mapped isometrically by $(\theta, \Phi)$ in~$\mathbb S^2$ (see Section~\ref{sec3}) and the separable coordinates coincide in this sector with the usual separable coordinates of $\mathbb S^2$. We have the same general non-globality of the coordinates whenever $k$ is not integer. The map $(\theta,\phi)$ has the same structure of the map $(r,\phi)$ of the case of the plane shown in Section~\ref{sec3}, as union of the maps $(\theta,\Phi)$, just imagine Figures~\ref{f2} and~\ref{f3} as views of the sphere from above the north pole.

\begin{figure}[t]
	\centering
	
	\includegraphics[width=4.5cm]{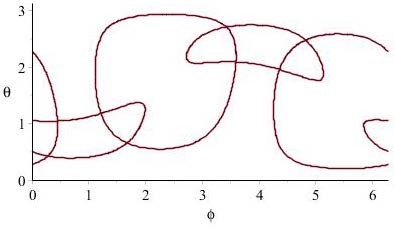}\qquad
	\includegraphics[width=4.5cm]{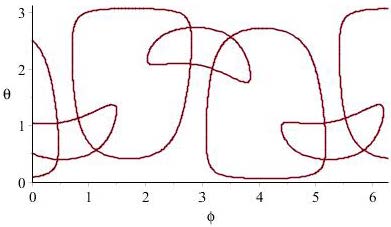}\qquad
	\includegraphics[width=4.5cm]{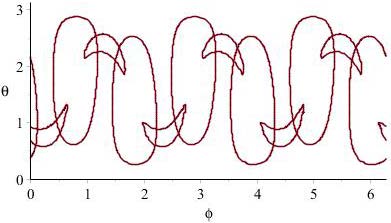}
	
\caption{Spherical-conical coordinates for $k=1$, $k=4/3$, $k=3$. The lines $\theta=0$ and $\theta =\pi$ coincide with the south and north poles of a sphere.}\label{f7}
\end{figure}

We plot below some separable coordinates of the manifold $(\mathbb S^2,G_k)$, see Figure~\ref{f7}. For $k=1$ the separable coordinates are the standard spherical-conical coordinates of $\mathbb S^2$, drawn as constant values of the eigenvalues of a Killing $2$-tensor of $(\mathbb S^2,G_k)$ for a given point. For $k=4/3$ we see that the separable coordinates are not global, since they are not $2\pi$ periodic in $\phi$.

The separable coordinates plotted above are the constant values of the eigenvalues of Killing tensors of $M_k$. When $k$ is not integer, it happens that these eigenvalues are not globally defined on the manifold, since, after a turn of $2\pi$ of $\phi$, they do not assume the same value. Since the components of the tensor are determined by the same eigenvalues, then also the tensors themselves are not globally defined, in these cases.

\begin{Remark}
 In all warped manifolds considered here, the ``warping" coordinates $(r,\Phi)$ are globally defined for any value of $k$, and they are the separable coordinates associated with the Killing tensor $\partial_\Phi \odot \partial_\Phi$.
\end{Remark}

It is easy to build Riemannian coverings for higher-dimensional spaces. For example, Let $(r,\theta,\phi)$ spherical coordinates on $\mathbb R^3$ and consider the metric $g_k$ such that
 \begin{equation*}
g_k\colon \ {\rm d}s^2=k^2\big({\rm d}r^2+r^2 \, {\rm d}\theta^2\big)+r^2\sin ^2(\theta) \, {\rm d}\phi^2.
 \end{equation*}
This metric is flat and it is locally equivalent to the Euclidean one. Let us consider now the coordinate transformation
\begin{gather*}
x=k r \sin(\theta) \cos \left(\frac \phi k \right),\qquad
y=k r \sin(\theta) \sin \left(\frac \phi k \right),\qquad
z=k r \cos (\theta).
\end{gather*}
This transformation pulls back the Euclidean metric into $g_k$ and it can be understood as a~covering map between $\mathbb E^3$ and its Riemannian coverings $(M_k,g_k)$ as done above for $\mathbb E^2$.

As a further example,
let us consider the unitary sphere $\mathbb S^3$ endowed with Hopf coordinates $(\eta,\xi_1 ,\xi_2)$ such that
\begin{eqnarray*}
	x=\cos \xi_1 \sin \eta, \qquad y=\sin \xi_1 \sin \eta,\qquad z=\cos \xi_2 \cos \eta, \qquad t=\sin \xi_2 \cos \eta,
\end{eqnarray*}
where $(x,y,z,t)$ are Cartesian coordinates of $\mathbb R^4$. The metric tensor of $\mathbb S^3$ is given by
 \begin{equation*}
{\rm d}s^2={\rm d}\eta^2+\sin ^2 \eta \, {\rm d}\xi_1^2+\cos^2 \eta \, {\rm d}\xi_2^2.
 \end{equation*}
By introducing parameters $a$, $b$ and $c$, we can write
\begin{equation*}
{\rm d}s^2=a^2 \, {\rm d}\eta^2+b^2 \sin ^2 \eta \, {\rm d}\xi_1^2+c^2 \cos^2 \eta \, {\rm d}\xi_2^2,
\end{equation*}
which is a metric
of constant curvature equal to $a^{-2}$. By proceeding in the same way as above, the metric can be understood as a metric on a Riemannian covering of $\mathbb S^3$.

We can fix $a$, which determines the curvature, and use $b$ and $c$ as parameters of a two-parameters covering of $\mathbb S^3$. The Riemannian covering map is consequently defined by
\begin{equation*}
\Xi_{(b,c)}(\eta, \xi_1,\xi_2) =(\eta, b\xi_1, c\xi_2).
\end{equation*}

\begin{Remark} An example of first integral of geodesics on coverings can be found in~\cite{M1,M2}. The system considered there is the geodesic flow on the Klein bottle, considered as double covering of the two-dimensional torus. It is shown that there exist a local first integral of the geodesic flow that is quadratic in the momenta, but not globally defined on the Klein bottle. The square of this local integral, instead, is globally defined, providing a global first integral of the geodesic flow of degree four in the momenta. The same behaviour can be observed among the first integrals of the geodesic flow on our Riemannian coverings of the plane, generated by the quadratic Killing tensors described above. Indeed, the Killing tensor (\ref{para}), together with its homologous obtained after the change $-\cos(k\phi) \rightarrow \sin(k\phi)$, $\sin(k\phi) \rightarrow \cos(k\phi)$, is globally defined only for integer values of $k$, but its square, whose components depend on $2k\phi$, is globally defined for any half-integer $k$, providing for these values a globally defined first integral of the geodesics of degree four (not the same for its cube). Similarly, the tensors (\ref{16}) and (\ref{17}) are globally defined for all half-integer values of $k$, but their squares are globally defined for $k=\frac n4$, where $n$ is any non-zero integer, since their components depend on $4k\phi$. This interesting fact should be studied more deeply, by considering the whole space of the Killing tensors and not only the squares of single tensors. We remark that another way of generating polynomial first integrals of the geodesics on the $M_k$ is provided by the extension procedure described below. It is enough to put the scalar potential equal to zero in the expressions for the first integrals of higher degree of the Tremblay--Turbiner--Winternitz, or Post--Winternitz, systems, in order to obtain first integrals of the geodesics, homogeneous in the momenta and of degree depending on~$k$. However, the behaviour of these integrals has not been studied yet.
\end{Remark}

\section{Superintegrability on covering manifolds}\label{sec5}

We study, through examples, how the superintegrability of Hamiltonian systems is affected by the Riemannian covering structure.

\begin{Definition} A $n$-degrees of freedom Hamiltonian system is \emph{Liouville integrable}, in short integrable, if it admits $n$ functionally independent first integrals all in Poisson involution.
It is \emph{superintegrable} if it admits more than $n$ functionally independent first integrals. It is said to be \emph{minimally superintegrable} if the first integrals are $n+1$, \emph{maximally superintegrable} if they are~${2n-1}$.
If all the first integrals are quadratic in the momenta, the Hamiltonian is \emph{quadratically integrable}, superintegrable, respectively.
\end{Definition}

Quadratically superintegrable systems are often characterized by separability of the Hamil\-ton--Jacobi equation in multiple coordinate systems (multiseparability).

\begin{Remark}
 We recall that the definition of superintegrability, as well as that of Liouville integrability, should take in account the existence of critical sets of zero measure, in the space phase, where functional independence of the first integrals fails, as in the following examples. We consider two one-dimensional harmonic oscillators with Hamiltonian
\[
H=\frac 12 \big(p_x^2 + p_y^2 +\omega_1^2 x^2 +\omega_2^2 y^2\big)
\]
with parameters $\omega_i$ having an irrational ratio, the sets of oscillations parallel to one of the Cartesian axes form two sets of orbits where the functional independence of the fist integrals fails and thus Arnold--Liouville theorem does not hold; indeed these orbits do not form a set diffeomorphic to $\mathbb T^2$, even if they correspond to level sets of the two first integrals in involution. Also for $\omega_1/\omega_2\in \mathbb Q $, when a third constant of motion is present, on these points of $ T^*\mathbb R^2$ the independence of the first integrals does not hold.
 Another classical example of superintegrable system where some points of the domain of the Hamiltonian has to be excluded is the Kepler motion on the plane. The first integrals that show Liouville integrability (the Hamiltonian and the angular momentum) are not everywhere functionally independent: their differential are linearly dependent on the points of $T^*\big(\mathbb R^2\setminus\{(0,0)\}\big)$ corresponding to circular orbits in the configuration space and, moreover, the solutions with null angular velocity are not complete, the corresponding motions are not periodic nor the associated orbits in the configuration space are closed curves (they are the motions falling into the origin in a finite time). Nevertheless, Hamiltonian systems where critical sets of the above type exist are considered superintegrable or integrable tout-court.
\end{Remark}

\subsection{The Tremblay--Turbiner--Winternitz system}

The Tremblay--Turbiner--Winternitz (TTW) Hamiltonian can be written as~\cite{TTWcdr,TTW}
\begin{equation}\label{TTW1}
H=\frac 12 p_r^2+\frac{1}{r^2}\left(\frac 12p_\Phi^2+\frac {\alpha_1}{\cos^2h\Phi}+\frac {\alpha_2}{\sin^2h\Phi}\right)+\omega r^2, \qquad r>0, \quad 0\leq \Phi <2\pi,
\end{equation}
where $\alpha_i$ and $\omega$ are real constants, $h$ is a positive rational number and $H$ is locally defined on the cotangent bundle of the Euclidean plane. This system is obtained in~\cite{TTWcdr}, where polar coordinates are denoted by $(u,q)$ instead of $(r,\phi)$, from the Hamiltonian
\begin{equation}\label{TTW2}
H=\frac 12p_r^2+\frac1{k^2r^2}\left(\frac 12p_\phi^2+\frac{c_1+c_2\cos \phi}{\sin^2\phi}\right)+\omega r^2,
\end{equation}
by putting $k=\frac 1{2h}$, $\Phi=k\phi$ and
 \begin{equation*}
c_1=\frac{\alpha_1+\alpha_2}{2h^2},\qquad c_2=\frac{\alpha_2-\alpha_1}{2h^2}.
 \end{equation*}
The metric tensor in (\ref{TTW2}) is the same as (\ref{wm}), and the Hamiltonian (\ref{TTW2}) is in the form of an extended Hamiltonian~\cite{TTWcdr}, for which polynomial first integrals of degree depending on $k\in \mathbb Q^+$ are explicitly determined via an algebraic-differential operator acting on
 \begin{equation*}
\frac 12p_\phi^2+\frac{c_1+c_2\cos \phi}{\sin^2\phi}.
 \end{equation*}
Therefore, we can consider the TTW system (\ref{TTW1}), which is equivalent to (\ref{TTW2}), as also equivalent~to
\begin{equation}\label{TTW3}
H=\frac 12p_r^2+\frac1{r^2}\left(\frac 12p_\Phi^2+\frac{\tilde c_1+\tilde c_2\cos \Phi}{\sin^2\Phi}\right)+\omega r^2,\qquad \Phi=k\phi,
\end{equation}
where $\tilde c_i=\frac {c_i}{k^2}$,
and we may consider (\ref{TTW2}) and (\ref{TTW3}) as defined on the manifold $M_k$, where (\ref{TTW2}) becomes (\ref{TTW3}) when $M_k$ is considered as the Riemannian covering of the Euclidean plane via the transformation $\Pi$, so that $0\leq \phi< 2\pi$. In this way we have a global Hamiltonian on $M_k$.


The TTW Hamiltonian admits polynomial first integrals of degree determined by $h$ for any positive rational $h$~\cite{TTWcdr, MPW, TTW}. The Hamiltonian (\ref{TTW1}) is globally defined, i.e., periodic in $\Phi$, only for $h$ integer or half-integer when $\alpha_i$ are not all zero, and this is the form from which the first integral of degree depending on $h$ is obtained in~\cite{MPW,TTW}. The global definition of $H$ and of its polynomial in the momenta first integrals is not considered in~\cite{MPW, TTW}. However, the lack of globality of (\ref{TTW1}) may not induce serious problems for some values of the parameter, since the particular form of the potential confines the dynamics between its singular values. Indeed, provided $\alpha_2\neq 0$, one can always choose non restrictive initial conditions such that the motion of the point is confined between $\Phi=0$ and $\Phi=\pi/2h$, if $\alpha_1\neq 0$, or $\Phi=\pi/h$ otherwise, so that, for any $h$, the half-line $\Phi=0$ at least prevents the existence of trajectories making circuits around the origin. Things are different when $\alpha_2=0$ and $\alpha_1 \neq 0$. In this case, when $h< 1/4$ the singularities disappear, the point can describe circuits around the origin and the periodicity in~$\Phi$ of $H$ and its first integrals becomes essential.
Summarizing the previous remarks, for $h<1/4$, the Hamiltonian (\ref{TTW1}) is no longer periodic in $\Phi$ and the system is not well defined and its peculiarities, such as the periodicity of the limited orbits are lost in the Euclidean plane.

 Conversely, in the form (\ref{TTW2}), i.e., when the system is defined on $M_k$, the superintegrability of $H$ can be proved independently from the fact that the rational parameter $k$ is integer or not. Indeed, the third first integral obtained by the extension procedure described in~\cite{TTWcdr} for any non-zero positive rational $k$ applies to the form (\ref{TTW2}) and involves trigonometric functions depending on $\phi$, that we can assume $0\leq \phi <2\pi$ (actually, we have the perfect coincidence with the (\ref{TTW1}) when $0\leq \phi < \frac 2k \pi$) and the problems of global definition described above do not arise.

 Therefore,

 \begin{Proposition} The Hamiltonian~\eqref{TTW1}, and its constants of motion, are globally defined on~$M_1$ only for $k=\frac 1{2h}$, with $h$ integer or half integer. Instead, \eqref{TTW2} and its constants of motion, obtained via the extension procedure, are well defined for any $k \in \mathbb Q^+$.
 \end{Proposition}

 When $\alpha_1=\alpha_2=0$, the TTW system becomes the harmonic oscillator with the metric $g_{k=\frac 1{2h}}$ and it can be considered as a system on the covering manifolds $M_{\frac 1{2h}}$. Now, the singularities of the general case do not confine the dynamics any more and the global definition of the first integrals becomes fundamental for the determination of integrability and superintegrability. We will consider these problems in the next example for the Kepler--Coulomb system determined on the $M_k$ as particular case of the Post--Winternitz system. Since the TTW and PW systems transform one into the other via coupling constant metamorphosis~\cite{PW}, what said below about the superintegrability of the KC system holds true for the harmonic oscillator also.

\subsection{The Kepler--Coulomb system}

Consider the parabolic coordinates of Section \ref{PC}
 \begin{equation*}
u=k^2\frac r{2}(\sin k \phi+1), \qquad v=k^2\frac r{2}(\sin k \phi-1),
 \end{equation*}
 \begin{equation*}
r=\frac{1}{k^2}(u-v), \qquad \Phi=k\phi=\arcsin \frac{u+v}{u-v}.
 \end{equation*}
In coordinates $(u,v)$, we have
\begin{equation*}
g_k=\frac{k^4}{u-v}\left(u \partial_u \odot \partial_u-v \partial_v \odot \partial_v\right), \qquad K=\frac{k^4 uv}{u-v}(\partial_u\odot \partial_u-\partial_v\odot \partial_v).
\end{equation*}
 The general separable scalar potential in $(u,v)$ is in St\"ackel form
 \begin{equation*}
V=g^{uu}\alpha(u)+g^{vv}\beta(v)=\frac {k^4}{(u-v)}(u \alpha-v\beta).
 \end{equation*}
If we choose
 \begin{equation*}
\alpha= \frac{a}{2k^2u},\qquad \beta=-\frac{a }{2k^2v},
 \end{equation*}
we have
 \begin{equation*}
V=\frac ar,
 \end{equation*}
that is, the Kepler--Coulomb (KC) potential on the covering manifold $M_k$, and $V$ is separable in both $(u,v)$ and $(r,\phi)$.
The potential $V$ is globally defined on any $M_k$, apart the singular point $r=0$.
The quadratic first integrals of the system are
\begin{gather}
H=\frac 12\biggl(p_r^2 +\frac 1{k^2r^2}p_\phi^2\biggr)+\frac ar, \label{KC} \\
L=\frac 1{2} p_\phi^2, \nonumber\\
K=\frac 12 \biggl(- k\cos k\phi\, p_r p_\phi+\frac{\sin k \phi}{r} p_\phi^2\biggr)+\frac a{2} \sin k \phi. \nonumber
 \end{gather}
The functions $H$, $L$, $K$ are all functionally independent and the Hamiltonian $H$ is therefore quadratically superintegrable, when all the integrals are globally defined. The first integral $K$ is associated to the generalisation to the covering space of the Laplace constant of motion of the Euclidean plane's Kepler--Coulomb system.
\begin{Proposition}
 The potential $V$ is multiseparable, and quadratically superintegrable, only for integer values of $k$.
\end{Proposition}

\begin{proof} The tensor $K$ is globally defined only for $k$ integer, while $H$ and $L$ are always globally defined. \end{proof}

Differently from what seen about the TTW system, in this case the orbits are not confined to sectors bounded by singular points and they can actually make circuits around the origin, wandering in all the covering space. The globality of first integrals becomes therefore essential for the superintegrability of the system.

\begin{Remark} The Kepler--Coulomb system in coordinates (\ref{KC}) can be solved via the Jacobi method for any real value of $k$ in exactly the same way as for $k=1$ in the Euclidean plane. Indeed, in the separated complete integral of the Hamilton--Jacobi equation $W=W_r+W_\phi$, the term $W_r$ is independent from $k$ and coincides with the corresponding term of the Euclidean case, while $W_\phi=\sqrt{2}kl\phi$, being $l$ the constant of motion determined by $L=l^2$. $W_\phi$ differs from the Euclidean case only for the factor $k$. No trigonometric functions are involved in the Jacobi method and all quantities depend only on $r$ and $\phi$, making the determination of the orbits well defined on the entire $M_k$ (apart the point $r=0$) for any real non-zero~$k$.
\end{Remark}

We recall now that the Kepler--Coulomb system (\ref{KC}) can be written as an instance of the Post--Winternitz (PW) system~\cite{PW}
\begin{equation}\label{powi}
H_{\rm PW}=p_r^2+\frac 1{r^2}\bigg( p_\phi^2+\frac 14f_2\bigg(\frac \phi 2\bigg)\bigg)-\frac Q{2r},\qquad f_2(x)=h^2\bigg( \frac \alpha{\cos ^2 hx}+\frac \beta{\sin ^2 hx}\bigg)
\end{equation}
for $\alpha=\beta=0$, where $(r,\phi)$ are assumed to be polar coordinates on the plane. The PW system can be deduced from the TTW system via coupling-constant-metamorphosis and it admits first integrals of degree depending on a rational parameter $h$, making it superintegrable for all positive rationals $h$. In~\cite{PW}, the high-degree first integrals of the system are obtained from (\ref{powi}) and they involve trigonometric functions of $h\phi$. Again, as for the TTW system, whenever $\alpha$ and $\beta$ are not zero, the singularities of the potential confine the dynamics in some part of the plane and the global definition of $H_{\rm PW}$ and of its first integrals is not an issue for $h$ non integer. When~$\alpha$,~$\beta$ are both zero, instead, the dynamics is much different and the globality becomes relevant.

The PW system and its high-order first integral can be formulated within the Hamiltonian extension procedure as described in~\cite{CDRPW}, in the form
\begin{equation}\label{PW1}
H=\frac 12 p_r^2+\frac {h^2}{4r^2}\left( \frac 12 p_\phi^2+\frac {c_1+c_2 \cos \phi}{\sin^2 \phi}\right)-\frac E{2r}.
\end{equation}
Here, we get the Kepler--Coulomb system for $c_1=c_2=0$, $E=-2a$. Therefore, the KC system remains in this case superintegrable for any positive rational value of the parameter $h$,
 and its superintegrability is now determined by a polynomial in the momenta first integral of degree depending on $h$, and not by the Laplace vector, which is equivalent to the quadratic first integral provided by $K$, globally defined only for $k=2/h$ integer. It follows that, for $h>2$ integer,~$K$~is not globally defined on $M_{\frac 1h}$. Therefore,

\begin{Proposition} The Hamiltonian~\eqref{powi}, and its constants of motion, are globally defined on~$M_1$ only for $k=\frac 1{2h}$, with $h$ integer or half integer. Instead, \eqref{PW1} and its constants of motion, obtained via the extension procedure, are well defined for any $k \in \mathbb Q^+$.
 \end{Proposition}

 \begin{proof} The system (\ref{PW1}) is not explicitly written on a covering manifold of $M_1=\mathbb E^2$ and its third first integral obtained in~\cite{CDRPW} is globally defined, since it depends on trigonometric functions of $\phi$. \end{proof}

\begin{Proposition} The KC system obtained from \eqref{PW1} is globally defined and Liouville integrable for all $k\in \mathbb R^+$.
For $k\in \mathbb N^+$, the system is quadratically superintegrable with the three first integrals $H$, $L$ and~$K$.
For $k\in \mathbb Q^+$, the system is superintegrable.
\end{Proposition}

\begin{proof} The first integrals $H$ and $L$ are well defined for any positive real value of $k$. For any positive integer $k$, the quadratic first integral $K$ is globally well defined, the system is therefore superintegrable. For any rational $k$ the Hamiltonian $H$ can be written as an extended Hamiltonian (as in (\ref{PW1}) with $h=2/k$ and $c_1=c_2=0$) so that the
 the Laplace constant of motion $K$ (which is not globally defined) is replaced by a polynomial in the momenta first integral of degree depending on $h$ obtained via the extension procedure.
Thus, the Hamiltonian is superintegrable on $M_k$ for all positive rational $k$.\end{proof}

An analogous discussion holds for the harmonic oscillator obtained as particular case of the TTW system.

\begin{Remark}
While the third first integral of the TTW and PW systems can be obtained by the extension procedure, avoiding problems of global definition, the expressions of the Killing tensors and vectors related to separability still involve trigonometric functions of $k\phi$, so that they are defined on the covering manifolds $M_k$ and fail globality for generic
non integer values of $k$. As instance the multiseparability in polar and parabolic coordinates of $H$ is lost if $k$ is not integer.
\end{Remark}

\begin{Remark}
The separation of the associated Laplace--Beltrami or Schr\"odinger operators on Riemannian coverings behaves exactly as in the standard case, as in~\cite{BCR-I, BCR-II}. Separability and superintegrability must take into account issues of global definition, similar to those exposed above for Hamilton--Jacobi separation and classical superintegrability. We remark that the procedure of extension applied to Schr\"odinger equation~\cite{ShiftLadder} does not involve trigonometric functions of $k\phi$, therefore, the symmetry operators obtained by that procedure are well defined on the mani\-fold~$(M_k,g_k)$ for any $k\in \mathbb Q^+$.
\end{Remark}

\section{Conclusions}\label{sec6}
We have now a better understanding of the metrics (\ref{wm}), appearing in~\cite{TTWcdr,MPW,TTW}. The metrics, and all the superintegrable systems of the Tremblay--Turbiner--Winternitz family on constant-curvature manifolds written in the form (\ref{TTW1}), are actually defined on the coverings $M_k$ of $M_1$ which corresponds to the Euclidean plane, or of the sphere or of the other constant-curvature manifolds. Some particular care must be taken when considering the global definition of the Hamiltonian systems on the $M_k$: the Hamiltonians, and its first integrals, should be actually periodic in $\Phi$. We have shown how the coverings $M_k$, for certain values of $k$, can be endowed with separability structures induced by those on $M_1$ and that, eventually, separable coordinate systems are defined on the same surface, simply by gluing together many copies of the separable coordinates of $M_1$. We have shown that superintegrability and separability on Riemannian coverings can depend on the value of $k$, since the Hamiltonian and the Killing tensors of different orders do not have necessarily all the same period in $\Phi$. As example we have shown that the quadratical superintegrability of the Kepler--Coulomb system on Riemannian coverings $M_k$ of the Euclidean plane holds for $k$ integer only. At the end of our analysis, still something not completely clear remains about metrics of the form (\ref{wm}). By introducing the transformations~$\Pi$, it is now evident that these metrics can be understood as metrics on Riemannian coverings of the Euclidean plane. On the other hand, by avoiding the use of $\Pi$ we can build structures that are apparently more general than those obtained otherwise, as it is shown in the examples of the superintegrability of the TTW, PW, KC and harmonic oscillator systems. However, as a matter of fact, we are unable to determine certain symmetries (Killing vectors and tensors) of (\ref{wm}) without the help of the map $\Pi$. We remark that the difference between integer and non-integer values of $k$ determines the fact that manifolds $M_k$ are true coverings of $M_1$ or not, as pointed out previously.

\subsection*{Acknowledgements}
The authors wish to thank Manuele Santoprete for useful discussions about the topic of this article, and the anonymous referees for their valuable suggestions.

\pdfbookmark[1]{References}{ref}
\LastPageEnding

\end{document}